\documentclass{PoS}
\usepackage{amsmath}
\usepackage{lineno}

\title{First data at Belle II and Dark Sector physics}

\ShortTitle{First data at Belle II and Dark Sector physics}

\author{\speaker{Giacomo De Pietro}%
        \thanks{On behalf of the Belle~II Collaboration.}\\
       Dipartimento di Matematica e Fisica, Universit\`a di Roma Tre and INFN Sezione di Roma Tre,\\Via della vasca navale 84, I-00146 Rome, Italy\\
       E-mail: \email{giacomo.depietro@roma3.infn.it}}

\abstract{Belle~II is a major upgrade of the Belle experiment and operates at the $B$-factory SuperKEKB in Japan. Since the SuperKEKB collider has a design luminosity of 8$\;\times\;$10$^{35}$~cm$^{-2}$s$^{-1}$, about 40 times larger than that of KEKB, Belle~II aims to collect $50$~ab$^{-1}$ of data over a period of 8 years.

The first data taking runs for physics analyses have started in April 2018, with a lower luminosity than the designed one for commissioning purposes. Even with the early dataset, having an integrated luminosity up to 20~fb$^{-1}$, Belle II can improve the current results in the Dark Sector field. In this paper we will present the expected sensitivity of Belle~II for invisibly decaying Dark Photons, for Axion-Like Particles and for invisibly decaying $Z^{\prime}$ assuming a $L_{\mu}-L_{\tau}$ model.}

\FullConference{The International Conference on B-Physics at Frontier Machines - BEAUTY2018\\
		6-11 May, 2018\\
		La Biodola, Elba Island, Italy}

\begin{document}

\section{The Belle~II experiment}
Belle~II is a $4\pi$ magnetic spectrometer~\cite{tdr} and is a major upgrade of the Belle experiment that operates at the $B$-factory SuperKEKB, located at the KEK laboratory in Tsukuba, Japan.

The SuperKEKB facility is designed to collide electrons and positrons at center-of-mass energies in the region of the $\Upsilon$ resonances. Most of the data will be collected at the $\Upsilon(4S)$ resonance ($\sqrt{s} =$ 10.58~GeV), which is just above threshold for $B$-meson pair production. In the case of $B\bar{B}$ production hence no additional fragmentation particles are produced. 
The collider is designed with asymmetric beam energies to provide a boost to the center-of-mass system and thereby allow for time-dependent $CP$ violation measurements. The boost is slightly lower than that at KEKB, which is advantageous for analyses with neutrinos and missing energy in the final state, that require a good detector hermeticity.
SuperKEKB has a design luminosity of 8$\;\times\;$10$^{35}$~cm$^{-2}$s$^{-1}$, about 40 times larger than KEKB, with the aim to collect $50$~ab$^{-1}$ of data over 8 years.

The first data taking runs for physics analyses have started in April 2018, with a lower luminosity than the designed one. This particular running condition (called Phase 2) serves mainly for machine commissioning and beam background studies, and we expect to reach the KEKB peak luminosity (10$^{34}$~cm$^{-2}$s$^{-1}$) and to collect up to 20~fb$^{-1}$ of data by the end of July 2018.

Since the Vertex Detector (VXD) in only partially installed during Phase 2, physics studies can rely on the Central Drift Chamber (CDC) for the charged tracks measurement, on the Time-Of-Propagation (TOP) and the Aerogel Ring-Imaging Cherenkov (ARICH) detectors for the charged particle identification, on the Electromagnetic Calorimeter (ECL) for the neutral clusters measurement and on the $K_L$ and Muon (KLM) detector for the long-living particles detection.



\section{Searches for Dark Photons}

A new vector particle $A^\mu$, belonging to a gauge group $U(1)$, can couple to the Standard Model (SM) electromagnetic current $J_{\text{SM}}^\mu$ via the so-called vector portal: $L \supset \epsilon A_{\mu}  J_{\text{SM}}^{\mu}$. The coupling $\epsilon = \kappa e /\cos\theta_W$ arises from kinetic mixing of the hypercharge $(Y)$ and the vector field strengths: $(\kappa/2)V_{\mu\nu}F_Y^{\mu\nu}$. In this case, $A$ is then often called a Dark Photon and it is denoted by $A^{\prime}$.

\begin{figure}[htb]
  \centerline{
    \includegraphics[width=0.41\textwidth]{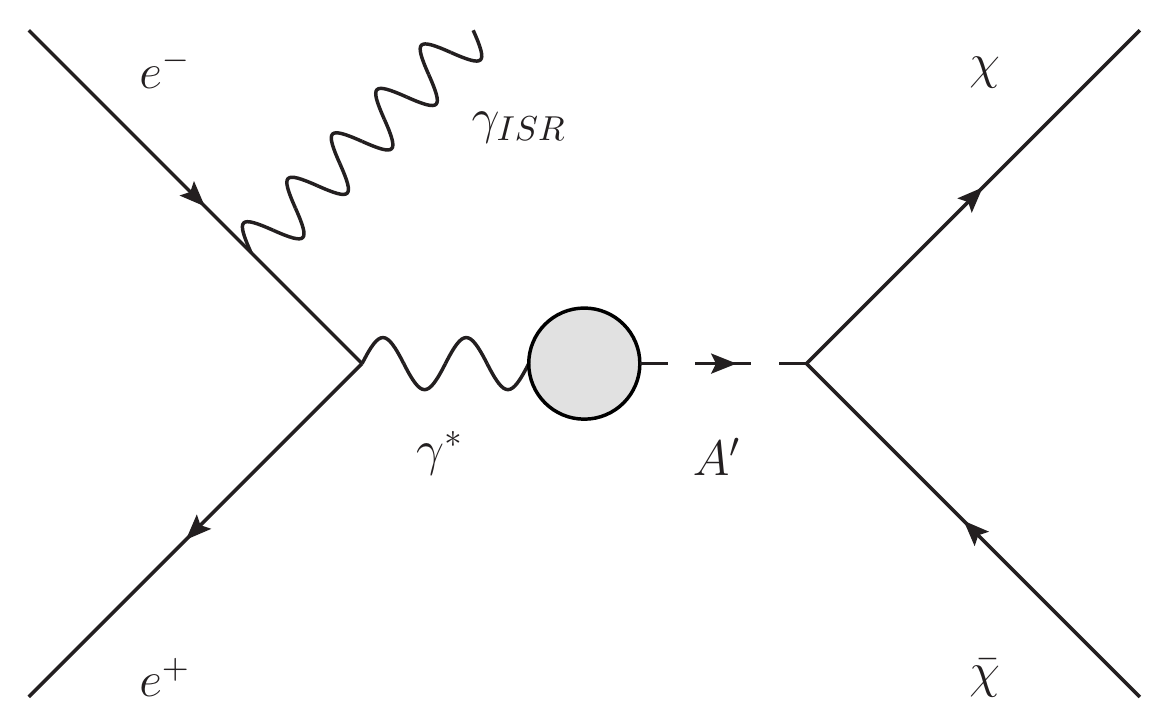}
  }
  \caption{Feynman diagram of the process $e^+e^- \to \gamma_{\text{ISR}} \; A^{\prime}$, with the subsequent decay $A^{\prime} \to \chi \bar{\chi}$.}
  \label{fig:feynman_darkPhoton}
\end{figure}

At Belle~II the Dark Photon can be searched for in the process $e^+ e^- \to \gamma_{\text{ISR}} \; A^{\prime}$, whose cross section is proportional to $\epsilon^2 \alpha / s$ (here $\alpha$ is the electromagnetic coupling). The Dark Photon can decay to SM final states $A^{\prime} \to l^+ l^-$ or $A^{\prime} \to h^+ h^-$ ($l$ = leptons, $h$ = hadrons), where branching fractions (BF) of the $A^{\prime}$ are the same as that of a virtual SM photon of mass $m_{A^{\prime}}$.
If $A^{\prime}$ is not the lightest Dark Matter (DM) particle, it will dominantly decay into light DM $\chi$ via $A^{\prime} \to \chi \bar{\chi}$. Since the DM particles do not interact with the detector, the experimental signature of this decay is a monochromatic photon ($\gamma_{\text{ISR}}$), having energy $E_{\gamma} = (s - m^2_{A^{\prime}}) \;/ \;2\sqrt{s}$, plus missing energy (Fig.~\ref{fig:feynman_darkPhoton}).

This search needs a dedicated first level (L1) trigger sensitive to single photons, that wasn't available at Belle and was only partially available at BaBar (BaBar recorded 53~fb$^{-1}$ of data with a single photon trigger, primarly at the $\Upsilon(2S)$ and $\Upsilon(3S)$ resonances~\cite{babar_darkPhotonInvisible}). Based on simulations including full beam backgrounds, for a trigger logic based on one ECL cluster with $E > 1$~GeV and the second most energetic ECL cluster below 300~MeV we expect a trigger rate of about 4 kHz in the barrel ECL and of about 7 kHz in the endcaps ECL. A second trigger logic will be used for this search (one ECL cluster with $E > 2$~GeV and a veto on Bhabba and $\gamma\gamma$ events), whose rate is expected to be about 5 kHz in the barrel ECL. These rates are mainly due to radiative Bhabba and $\gamma\gamma$ events in which only a single photon is produced within the detector acceptance.

A full detector simulation, including all the relevant QED backgrounds, was performed in order to evaluate the sensitivity of Belle~II (also during the Phase~II running condition) to $A^{\prime}$ decaying into an invisible final state~\cite{b2tip}. The main sources of background for this search have been found to be radiative Bhabba and $\gamma\gamma$ events where all but one photon are not detected by Belle~II, mainly because of small but not-neglibile photon detection inefficiencies in the ECL (the contribution of the irreducible background from $e^+ e^- \to \nu \bar{\nu} \gamma$ is negligible due to the smaller cross section of the process).

\begin{figure}[htb]
  \centerline{
    \includegraphics[width=0.76\textwidth]{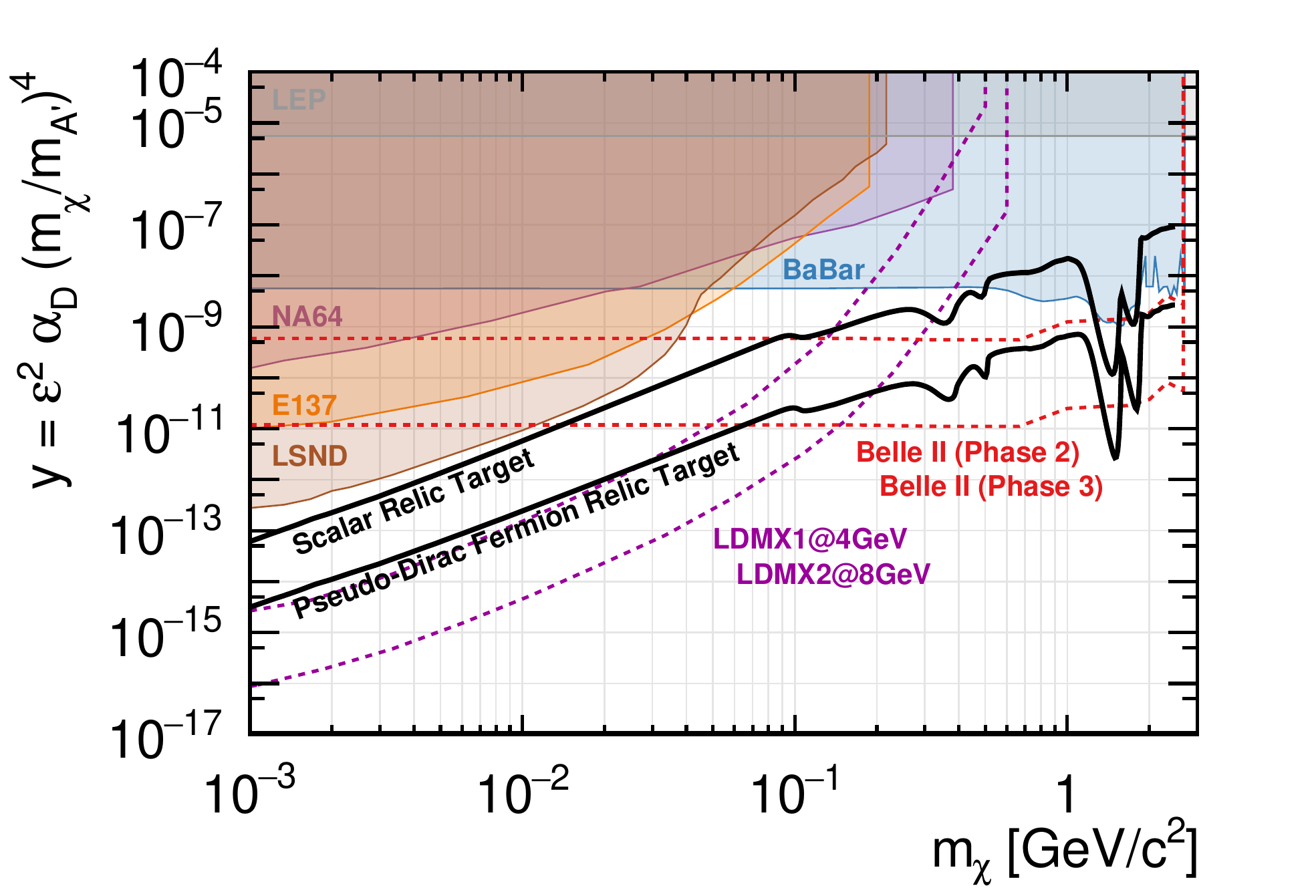}
  }
  \caption{Expected upper limits (90\% CL) on $y = \epsilon^2 \alpha_{\text{D}}$ (where $\alpha_D$ is the coupling with the dark sector) for the process $e^+e^- \to A^{\prime}$, $A^{\prime} \to \chi \bar{\chi}$ for a 20~fb$^{-1}$ dataset (Phase~2) and for a 50~ab$^{-1}$ dataset (Phase~3, full dataset). In this plot, $\alpha_{\text{D}} = 0.5$ and $m_{\chi} = \frac{1}{3}\; m_{A^{\prime}}$ are assumed. Figure taken from \cite{b2tip}.}
  \label{fig:darkPhotonInvisible}
\end{figure}

The expected Belle~II sensitivity for an integrated luminosity of 20~fb$^{-1}$ is shown in Fig.~\ref{fig:darkPhotonInvisible}. The better expected sensitivity compared to BaBar is due to the more homogeneous electromagnetic calorimeter of Belle~II, whose barrel part has no projective gaps to the interaction point. The expected sensitivity with the full dataset of 50~ab$^{-1}$ is also shown. 

\section{Searches for Axion-Like Particles}

Axion-Like Particles (ALPs) $a$ are hypothetical pseudo-scalar particles that can couple to the SM gauge bosons via the so-called axion portal. Axions were originally motivated by the strong $CP$ problem and have a fixed relation between coupling strength and mass. While the Axion and its parameters are related to QCD, the coupling and mass of ALPs are taken to be independent and can appear in a variety of extensions to the SM. The simplest search for an ALP at Belle~II is via its coupling to photons: $L \supset - \frac{g_{a\gamma\gamma}}{4} a F_{\mu\nu} \tilde{F}^{\mu\nu}$ (where $\tilde{F}^{\mu\nu} = \frac{1}{2} \epsilon^{\mu\nu\rho\sigma}F_{\rho\sigma}$).

\begin{figure}[htb]
  \centerline{
    \includegraphics[width=0.44\textwidth]{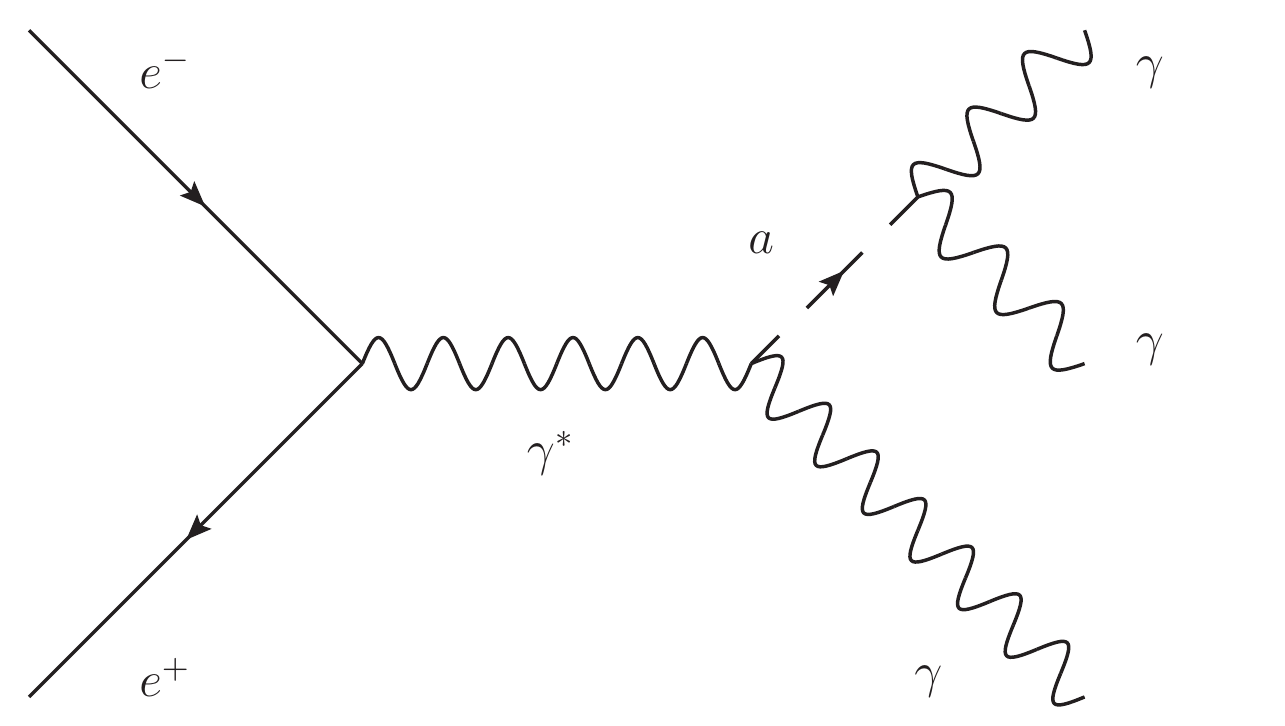}
  }
  \caption{Feynman diagram of the process $e^+e^- \to \gamma \; a$ (ALP-strahlung production), with the subsequent decay $a \to \gamma \gamma$.}
  \label{fig:feynman_Alp}
\end{figure}

There are two different production processes of interest at Belle~II: ALP-strahlung ($e^+e^- \to \gamma \; a$) and photon fusion ($e^+e^- \to e^+e^-a$). Even if ALP production via photon fusion typically dominates over ALP-strahlung (unless $m_a$ is close to $\sqrt{s}$), the final state in photon fusion production features only two soft photons (from $a \to \gamma\gamma$) and missing momentum which will lead to very high QED background. The most promising search is therefore from ALP-strahlung production (Fig.~\ref{fig:feynman_Alp})~\cite{torben}.

The decay width $\Gamma_a$ of the ALP is given by $\Gamma_a = \frac{1}{64\pi} \; g_{a\gamma\gamma}^2 \; m_a^3$. If $g_{a\gamma\gamma} \ll 1$~TeV$^{-1}$ and $m_a \ll 1$~GeV, it is important to notice that the decay width is particularly small and therefore the decay length can be very large, especially if the the ALP is significantly boosted. In such a case, the ALPs can escape from the detector region before decaying and the experimental signature is the same as for an invisibly decaying Dark Photon (only a single photon in the detector).

Depending on the values of mass and coupling, the signature of visible ALP decays can differ substantially. If the ALP has an high mass, it is produced with a small boost, so the opening angle of the decay photons is large and the final state is given by three resolved and detectable photons. If $m_a \lesssim 400$~MeV, the ALP is produced with a large boost and the opening angle between the two decay photons can not be resolved in the L1~trigger. In such a case, these events appear as $\gamma\gamma$ events and it is mandatory that they are not vetoed or pre-scaled at L1~trigger level, but passed to the offline reconstruction in order to maintain a high efficiency for low masses.  

The main background component is given by the QED process $e^+e^- \to \gamma\gamma\gamma$. A smaller component (especially for low masses) arises from $e^+e^- \to \gamma\gamma$ processes with a third photon coming from the beam background, but this background is expected to be significantly reduced by the good time resolution of the ECL (of the order of few nanoseconds).

\begin{figure}[htb]
  \centerline{
    \includegraphics[width=0.48\textwidth]{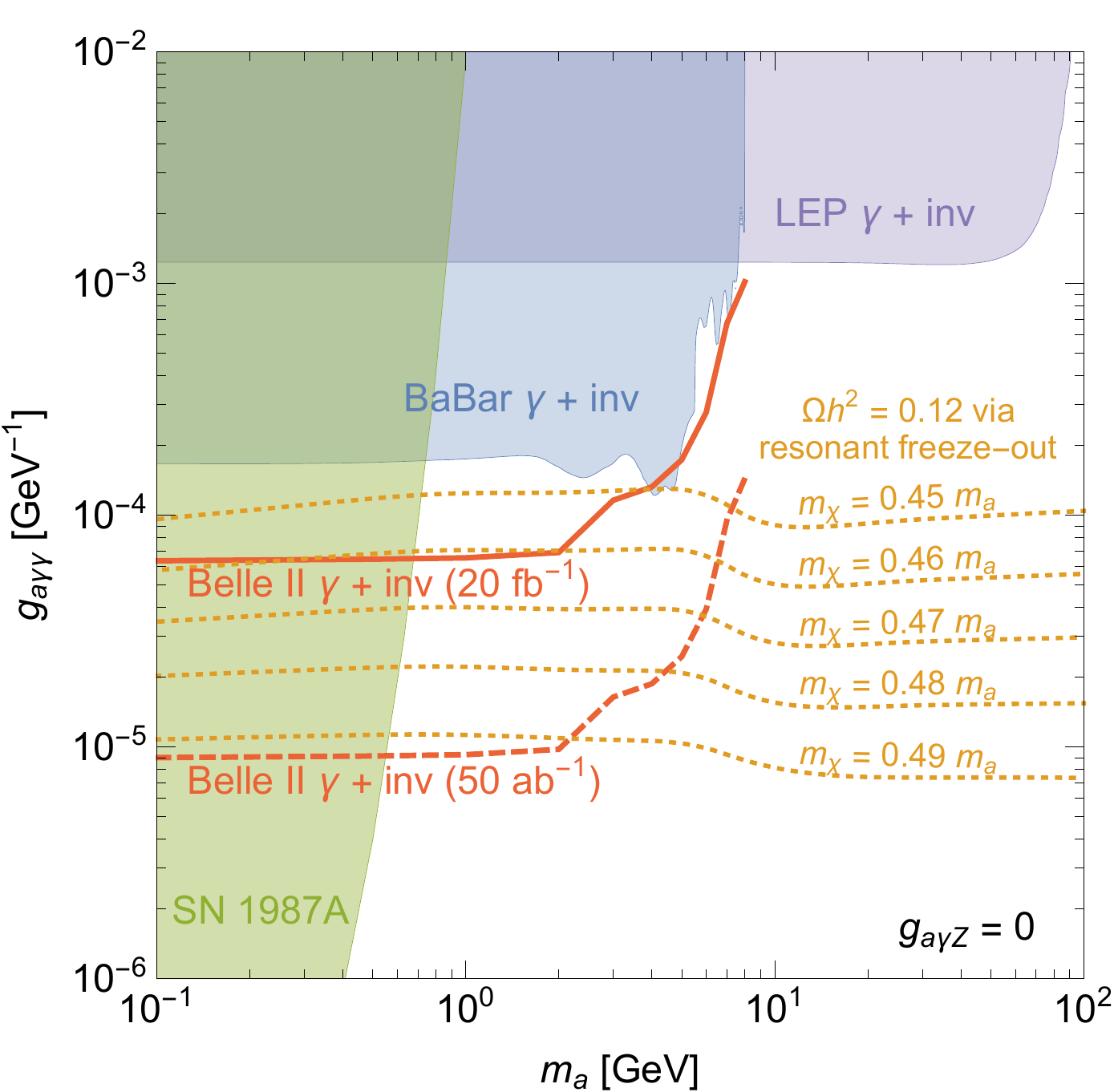}\qquad
    \includegraphics[width=0.48\textwidth]{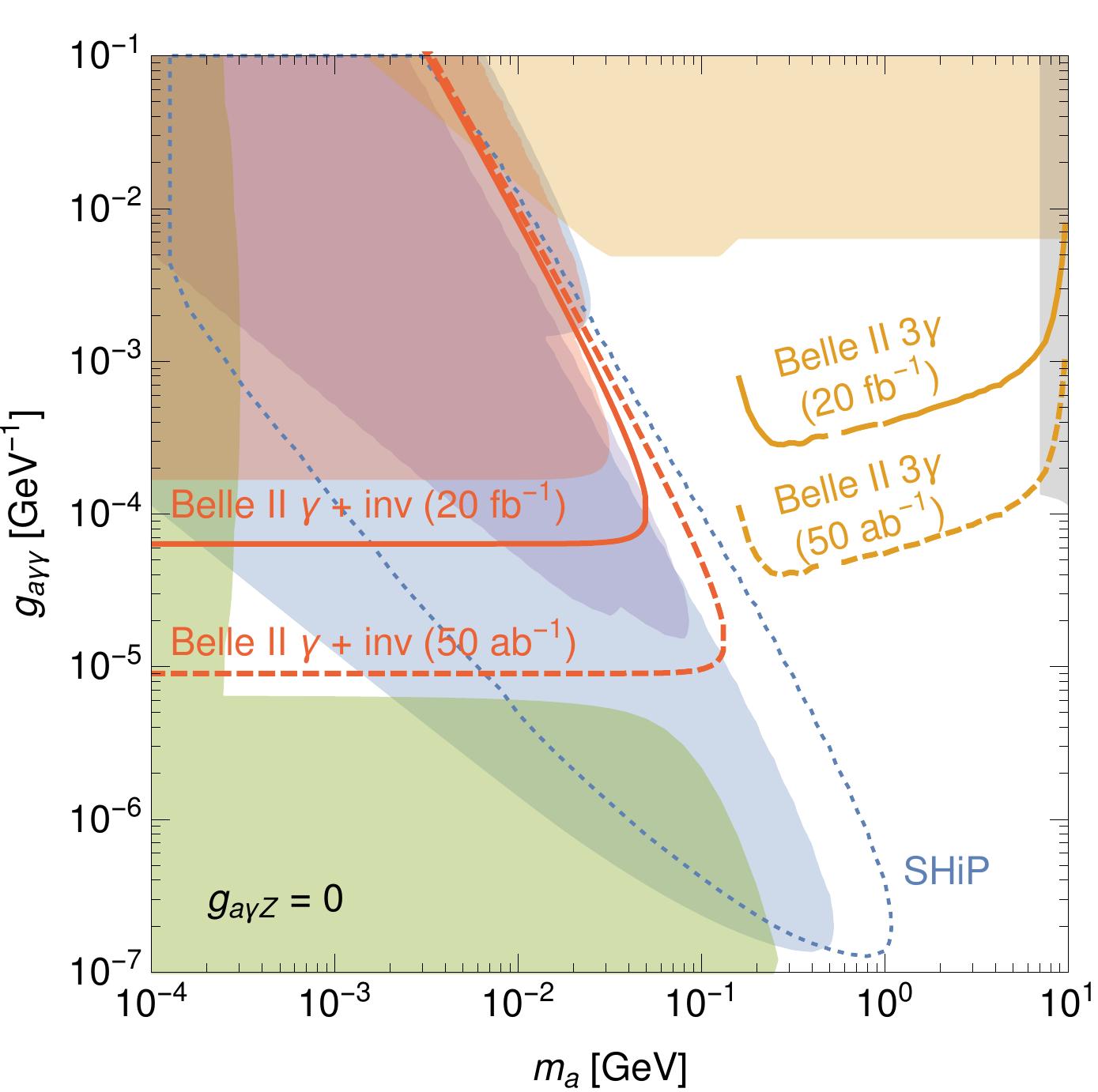}
  }
  \caption{Expected upper limits (90\% CL) on $g_{a\gamma\gamma}$ for the process $e^+e^- \to \gamma \; a$ for a 20~fb$^{-1}$ dataset (Phase~2) and for a 50~ab$^{-1}$ dataset (Phase~3, full dataset). On the left: limits for $a \to \chi \bar{\chi}$ compared to the parameter
region where one can reproduce the observed DM relic abundance via resonant annihilation of DM into photons; on the right: limits for $a \to \gamma \gamma$ (where the coupling $g_{a\gamma Z}$ is assumed to be much smaller than $g_{a\gamma\gamma}$) compared to the expected sensitivity from SHiP. Figures taken from~\cite{torben}.}
  \label{fig:AlpLimits}
\end{figure}

The expected Belle~II sensitivity for the process $e^+e^- \to \gamma \; a \; (\to \gamma\gamma)$ is shown in Fig.~\ref{fig:AlpLimits} considering an integrated luminosity of 20~fb$^{-1}$ (Phase 2) and 50~ab$^{-1}$ (Phase 3, full Belle~II dataset). In the same figure we show also the sensitivity for the case in which the ALP decays into light Dark Matter particles $\chi$ via $a \to \chi \bar{\chi}$ in ALP-strahlung processes (if $m_{\chi} \leq \frac{1}{2}\; m_a$). In such a case, the trigger and selection criteria considered are similar to the search for an invisibly decaying Dark Photon.

\section{Searches for $Z^{\prime}$ ($L_{\mu}-L_{\tau}$ model)}

It is possible to consider other extensions of the SM, in which a new massive vector boson $Z^{\prime}$, with a mass in the range $m_{Z^{\prime}} \sim$~MeV - GeV and a coupling to SM $g^{\prime} \sim 10^{-6}$~-~$10^{-2}$, is coupled to the $L_{\mu}-L_{\tau}$ current~\cite{zprime}. In this scenario, the $Z^{\prime}$ can not interact directly with the light leptons (both $e$ and $\nu_e$). An interesting case to study (not covered by Belle and BaBar) is the one in which the $Z^{\prime}$ decays into neutrinos ($\nu_{\mu}$ or $\nu_{\tau}$) or light DM particles, so into an invisible final state. Even if the decay into DM is kinematically forbidden, the $Z^{\prime}$ still has a sizeable decay width into two neutrinos\footnote{In general, the decay width to one neutrino species is half of the decay width to one charged lepton flavour. The reason is, of course, that the $Z^{\prime}$ only couples to left-handed neutrino chiralities whereas it couples to both left-handed and right-handed charged leptons.}: in the ``worst case scenario'' ($m_{Z^{\prime}} \geq 2\;m_{\tau}$) we have BF$(Z^{\prime} \to \nu\bar{\nu}) \sim 1/3$.

\begin{figure}[htb]
  \centerline{
    \includegraphics[width=0.44\textwidth]{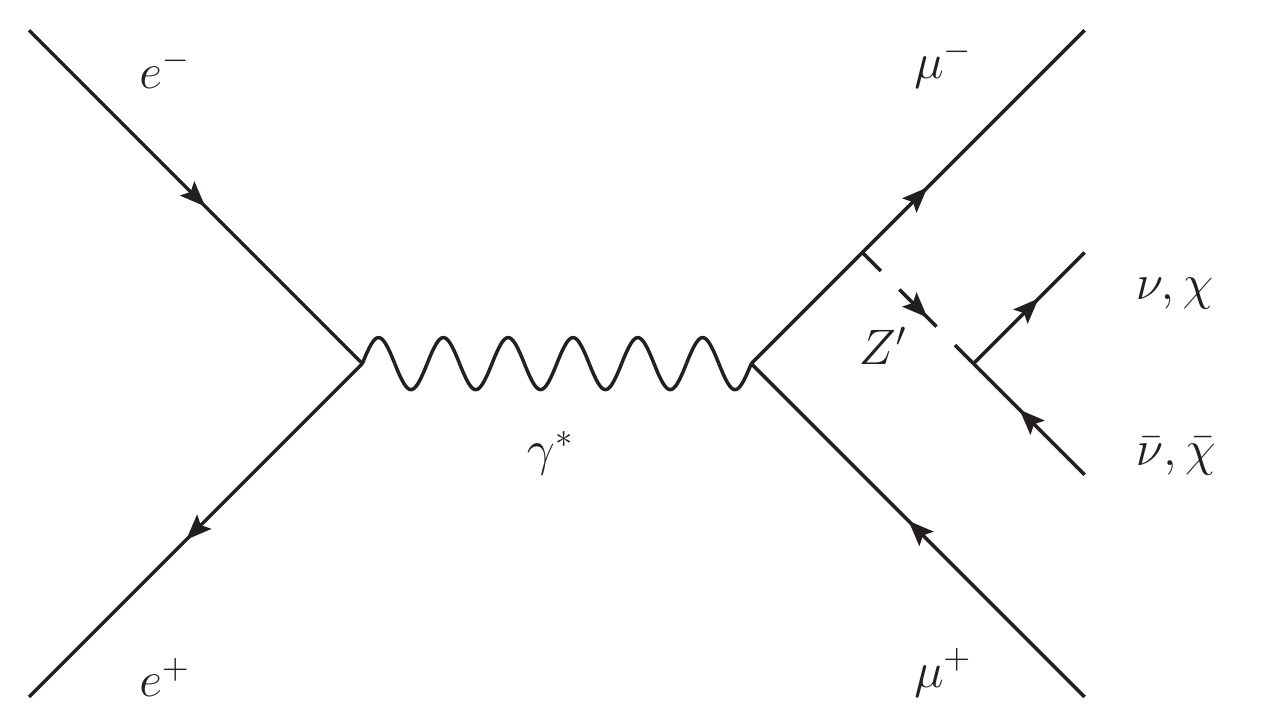}
  }
  \caption{Feynman diagram of the process $e^+e^- \to \mu^+ \mu^- \; Z^{\prime}$, with the subsequent decay $Z^{\prime} \to \nu \bar{\nu}$ or $Z^{\prime} \to \chi \bar{\chi}$.}
  \label{fig:feynman_Zprime}
\end{figure}

In principle, it is possible to search for a $Z^{\prime}$ produced with a pair of $\tau$ leptons ($e^+e^- \to \tau^+\tau^-\;Z^{\prime}$), but due to the experimental difficulties to study this particular process, we focused only to the $Z^{\prime}$ production accompanied by muons (Fig.~\ref{fig:feynman_Zprime}). In such a case, the experimental signature is given by two muons in the detector acceptance plus missing energy, and we look for a peak in the spectrum of the recoil mass with respect to the two muons. It is important to notice that, for this process, the cross section is proportional to $g^{\prime 2}$ and it goes to zero when $m_{Z^{\prime}}$ approaches $\sqrt{s}$.

We performed a full detector simulation in order to evaluate the sensitivity of Belle~II to the $Z^{\prime}$ decaying into an invisible final state.
In the simulation we took into account all the main SM backgrounds in which there can be only two muons in the final state: $e^+e^- \to \mu^+\mu^-(\gamma)$ (relevant for $m_{Z^{\prime}} \lesssim$ 2~GeV), $e^+e^- \to \tau^+\tau^-(\gamma)$ (2 $\lesssim m_{Z^{\prime}} \lesssim$ 7~GeV) and $e^+e^- \to e^+ e^- \mu^+\mu^-$ ($m_{Z^{\prime}} \gtrsim$ 7~GeV). To reject the background, we select two muon candidates and impose the following requirements: there are only two tracks in the event coming from the interaction point; the $p_T$ of the dimuon candidate is larger than 1~GeV; the recoil momentum points to the barrel ECL with no photons within a 15$^{\circ}$ cone around it; the extra energy in the ECL is lower than 1 GeV. Including a L1~trigger efficiency of $\sim 85\%$, the total signal efficiency with this selection is $\sim 35\%$\footnote{Since the selection criteria presented here are not optmized, the final selection that will be used for the analysis can differ.}.

\begin{figure}[htb]
  \centerline{
    \includegraphics[width=0.75\textwidth]{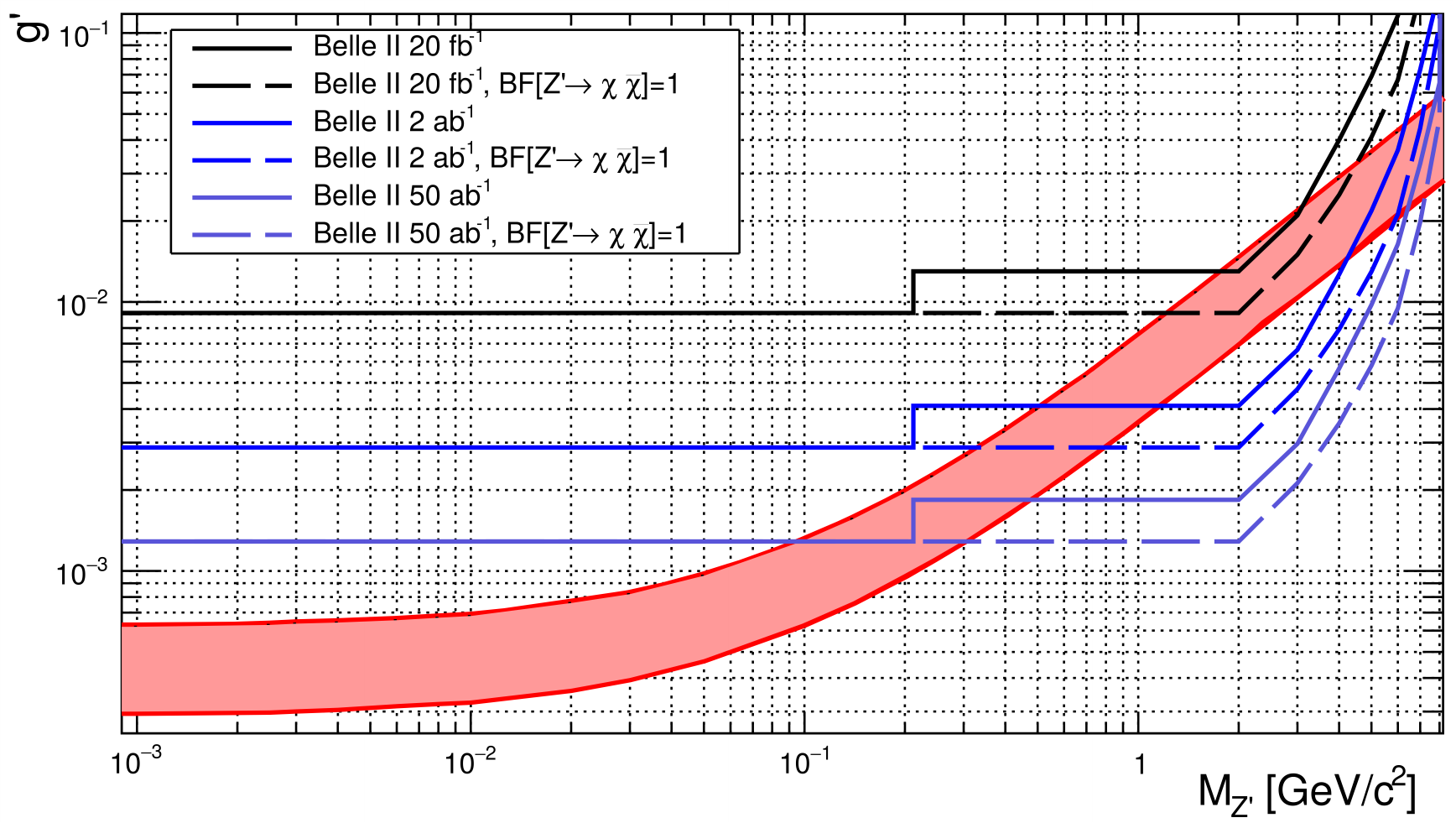}
  }
  \caption{Expected upper limits (90\% CL) on $g^{\prime}$ for the process $e^+e^- \to \mu^+ \mu^- \; Z^{\prime}$, $Z^{\prime} \to \nu \bar{\nu}$ or $Z^{\prime} \to \chi \bar{\chi}$ for a 20~fb$^{-1}$ dataset (Phase~2), for a 2~ab$^{-1}$ dataset (Phase~3, early dataset) and for a 50~ab$^{-1}$ dataset (Phase~3, full dataset). The dashed lines refer to the case in which the $Z^{\prime}$ decays to DM with BF$(Z^{\prime} \to \chi\bar{\chi})=1$. The red band refers to set of values that could explain the anomalous magnetic moment of the muon $\left((g-2)_{\mu} \pm 2\sigma\right)$.}
  \label{fig:Zprime}
\end{figure}

In Fig.~\ref{fig:Zprime} the solid histograms show the expected sensitivity of the Belle II~experiment to invisible decays of the $Z^{\prime}$ into neutrinos, whereas the dashed histograms show the expected sensitivity under the assumption that the invisible $Z^{\prime}$ branching fraction can be enhanced, up to BF$(Z^{\prime} \to \text{invisible}) = 1$, by the existence of light DM. We show both the sensitivities for different integrated luminosities: 20~fb$^{-1}$ (Phase~2), 2~ab$^{-1}$ (early Phase~3) and 50~ab$^{-1}$ (Phase~3, full Belle~II dataset). In the two latter cases, we scaled the expected background from the Phase~2 case accordingly with the integrated luminosity.


\section{Conclusions}

The Belle~II experiment has started the first data taking runs for physics analyses in April 2018. Thanks to an hermetic and upgraded detector, dedicated trigger logics and an improved offline reconstruction, Belle~II will give the world leading sensitivity for the presented Dark Sector searches even with a small dataset compared to the previous generation of $B$-factories.

In addition to the presented searches, Belle~II has a wide physics program related to the Dark Sector field~\cite{b2tip}, covering several topics such as visible decays of Dark Photons, displaced vertices of long-lived particles, decays of a Dark Higgs and Dark Scalars, Lepton Flavour Violating processes related to Dark Bosons, invisible decays of $\Upsilon$ resonances, magnetic monopoles, etc.

\end{document}